\documentclass[aps,pra,twocolumn,superscriptaddress]{revtex4-1}
\usepackage{graphicx}
\usepackage{amsmath}

\newcommand{\figref}[1]{Fig.~\ref{fig:#1}}
\newcommand{\equref}[1]{Eq.~(\ref{eq:#1})}
\newcommand{\tabref}[1]{Table~\ref{tab:#1}}
\newcommand{\secref}[1]{Sec.~\ref{sec:#1}}
\begin{document}


\title{Crossover trimers connecting continuous and discrete scaling regimes}


\author{Shimpei Endo}
\email[E-mail address: ]{endo@cat.phys.s.u-tokyo.ac.jp}
\affiliation{Department of Physics, University of Tokyo, 7-3-1 Hongo, Bunkyo-ku, Tokyo 113-0033, Japan}
\author{Pascal Naidon}
\affiliation{RIKEN Nishina Center, Saitama 351-0198, Japan}
\affiliation{ERATO Macroscopic Quantum Control Project, JST, 7-3-1 Hongo, Bunkyo-ku, Tokyo 113-0033, Japan}
\author{Masahito Ueda}
\affiliation{Department of Physics, University of Tokyo, 7-3-1 Hongo, Bunkyo-ku, Tokyo 113-0033, Japan}
\affiliation{ERATO Macroscopic Quantum Control Project, JST, 7-3-1 Hongo, Bunkyo-ku, Tokyo 113-0033, Japan}



\date{\today}

\begin{abstract}
For a system of two identical fermions and one distinguishable particle interacting via a short-range potential with a large $s$-wave scattering length, the Efimov trimers and Kartavtsev-Malykh trimers exist in different regimes of the mass ratio. The Efimov trimers are known to exhibit a discrete scaling invariance, while the Kartavtsev-Malykh trimers feature a continuous one. We point out that a third type of trimers, ``crossover trimers", exist universally regardless of short-range details of the potential. These crossover trimers have neither discrete nor continuous scaling invariance. We show that the crossover trimers continuously connect the discrete and continuous scaling regimes as the mass ratio and the scattering length are varied. We identify the regions for the Kartavtsev-Malykh trimers, Efimov trimers, crossover trimers, and non-universal trimers in terms of the mass ratio and the $s$-wave scattering length by investigating the scaling property and universality (i.e. independence on short-range details) of the trimers.
\end{abstract}


\maketitle

\section{Introduction}

Universality and the scaling symmetry are among the central concepts in physics. While many of the well-known examples appear in many-body systems, few-body systems can also show universality and possess exotic scaling features. A prime example is the Efimov states \cite{efimov1973energy,braaten2006universality,kraemer2006evidence,PhysicsF,ferlaino2011efimov}. The Efimov states are universal three-body bound states which exist for a three-particle system with short-range interactions when the $s$-wave scattering length $a_s$ between particles is very large. Their possibility has been discussed in various kinds of systems, including nucleons \cite{AnnRev_HamPlatt}, nuclear halo systems \cite{PhysRevLett.73.2817,nakamura2006observation,nakamura2009halo}, three-heliums system \cite{PhysRevA.86.012502}. Recently, the first experimental signature of the Efimov states has been observed in the ultracold atomic gases, where the large scattering length can be realized with the use of Feshbach resonances \cite{kraemer2006evidence,PhysicsF}.

\begin{figure}[b]
 \includegraphics[width=9cm,clip]{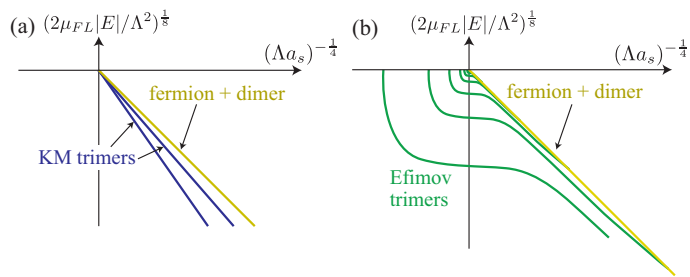}%
 \caption{\label{fig:Schematic_UniEfi_Difference}(Color online) Schematic illustration of the energy spectrum for (a) Kartavtsev-Malykh (KM) trimers and (b) Efimov trimers, where $\mu_{FL}=\frac{m_F m_L}{m_L + m_F}$, $a_s$ is the $s$-wave scattering length, and $\Lambda$ is the three-body parameter. The rightmost lines show the dimer binding energy. }
 \end{figure}

The Efimov states are known to have two important features \cite{efimov1973energy,braaten2006universality}. One of them is the discrete scaling invariance: if one knows the energy level of one Efimov state, all the other energy levels scale by a universal scaling factor, as illustrated in \figref{Schematic_UniEfi_Difference} (b). This discrete scaling invariance originates from the universal inverse square three-body attraction at a hyperradius $R$ smaller than the $s$-wave scattering length (see \cite{braaten2006universality} for the definition of hyperradius). This Efimov attraction has another important consequence known as the Thomas effect \cite{Thomas_effect}: under this attraction, particles tend to come closer all the way down to an infinitesimally small distance, which implies the collapse of the system. In a real system, however, the interaction between particles has a finite range with a repulsive core, which prevents such a collapse. One can take into account this finite-range nature of the potential by imposing a three-body boundary condition at short distance, which can be expressed in terms of a single three-body parameter $\Lambda$.  Thus, the Efimov states are characterized by two parameters: the $s$-wave scattering length $a_s$, and the three-body parameter $\Lambda$. Except for these parameters, all the other details of the inter-particle potentials plays negligible role when the $s$-wave scattering length is large, and the Efimov trimers can be described in a universal (i.e., model-independent) manner.

\begin{figure*}[t]
\includegraphics[width=18cm,clip]{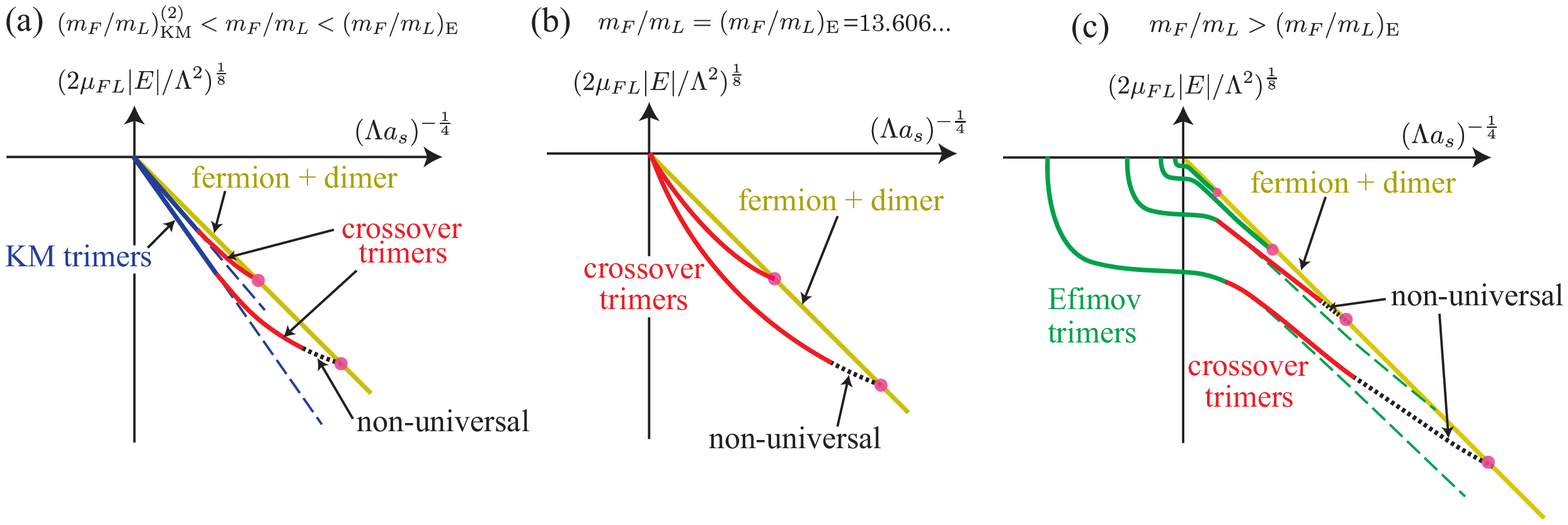}%
\caption{\label{fig:Result_Spectrum_Sche}(Color online) (a)-(c) Schematic illustrations of the energy spectra for different regimes of the mass ratio. Curves are deformed from the real ones for the sake of illustration. The crossover trimer region (red curve), Efimov trimer region (green curve), Kartavtsev-Malykh (KM) trimer region (blue curve), and non-universal region (black dotted curve) are distinguished according to their scaling properties (see \secref{result} for their identification). The dashed curves in (a) and (c) are the ones as predicted from the continuous scaling law of the KM trimers and the discrete scaling law of the Efimov trimers, respectively. The dots on the positive $a_s$ side indicate the fermion-dimer dissociation points at which the trimer dissociates into a fermion and a dimer and the fermion-dimer $p$-wave scattering volume diverges. In Fig.~(c), only four levels of the Efimov series are shown.}
\end{figure*}

Whether the Efimov states exist or not depends on the statistics, mass ratios, and $s$-wave scattering lengths of the particles \cite{efimov1973energy}. In this paper, we consider a system with two identical spin-polarized fermions (i.e. fermions without internal degree of freedom) and one distinguishable particle that interact with a large $s$-wave scattering length $a_s$, i.e. close to the so-called unitarity limit corresponding to $1/a_s =0$. We assume that there is no interaction between the fermions, since the $s$-wave interaction between identical fermions is forbidden by the Pauli principle, and higher partial-wave interactions are well suppressed at low energy. The Fermi-Dirac statistics dictates that the angular momentum of the trimer states be non-zero, so that the centrifugal repulsion competes with the Efimov attraction. The centrifugal repulsion becomes weaker for heavier fermions, and the Efimov states are known to exist if and only if the mass ratio between the fermion and the other particle $m_F/m_L$ is larger than $(m_F/m_L)_{\mathrm{E}}=13.606...$ \cite{efimov1973energy,PhysRevA.67.010703}. Below this critical mass ratio, the centrifugal barrier prevails and there emerges an $R^{-2}$ effective repulsive potential between the three particles at short distance $R\ll |a_s|$. Nevertheless, it has recently been shown by Kartavtsev and Malykh \cite{kartavtsev2007low} that such two fermions plus one particle system can still support universal three-body bound states below the critical mass ratio in the limit of the large $s$-wave scattering length. In fact, a first trimer state emerges for the mass ratio $m_F/m_L>(m_F/m_L)_{\mathrm{KM}}^{(1)}=8.172...$, and a second one appears for $m_F/m_L>(m_F/m_L)_{\mathrm{KM}}^{(2)}=12.917...$. These Kartavtsev-Malykh trimers (KM trimers) have properties distinct from the Efimov states. Since the potential is repulsive at short distance, the three particles cannot come close. Thus, these trimer states no longer depend on $\Lambda$, and they are characterized only by $a_s$. As a consequence, their spectrum exhibits a continuous scaling invariance: if one finds one of such trimers at a specific value of the $s$-wave scattering length, properties of trimers for different values of the $s$-wave scattering length can be predicted by continuous rescaling of the $s$-wave scattering length and the energy as $a_s \rightarrow \beta a_s$, $E\rightarrow \beta^{-2}E$, and $\langle r \rangle \rightarrow \beta \langle r \rangle$ (see \figref{Schematic_UniEfi_Difference}~(a)).

\begin{figure*}[t]
\includegraphics[width=18cm,clip]{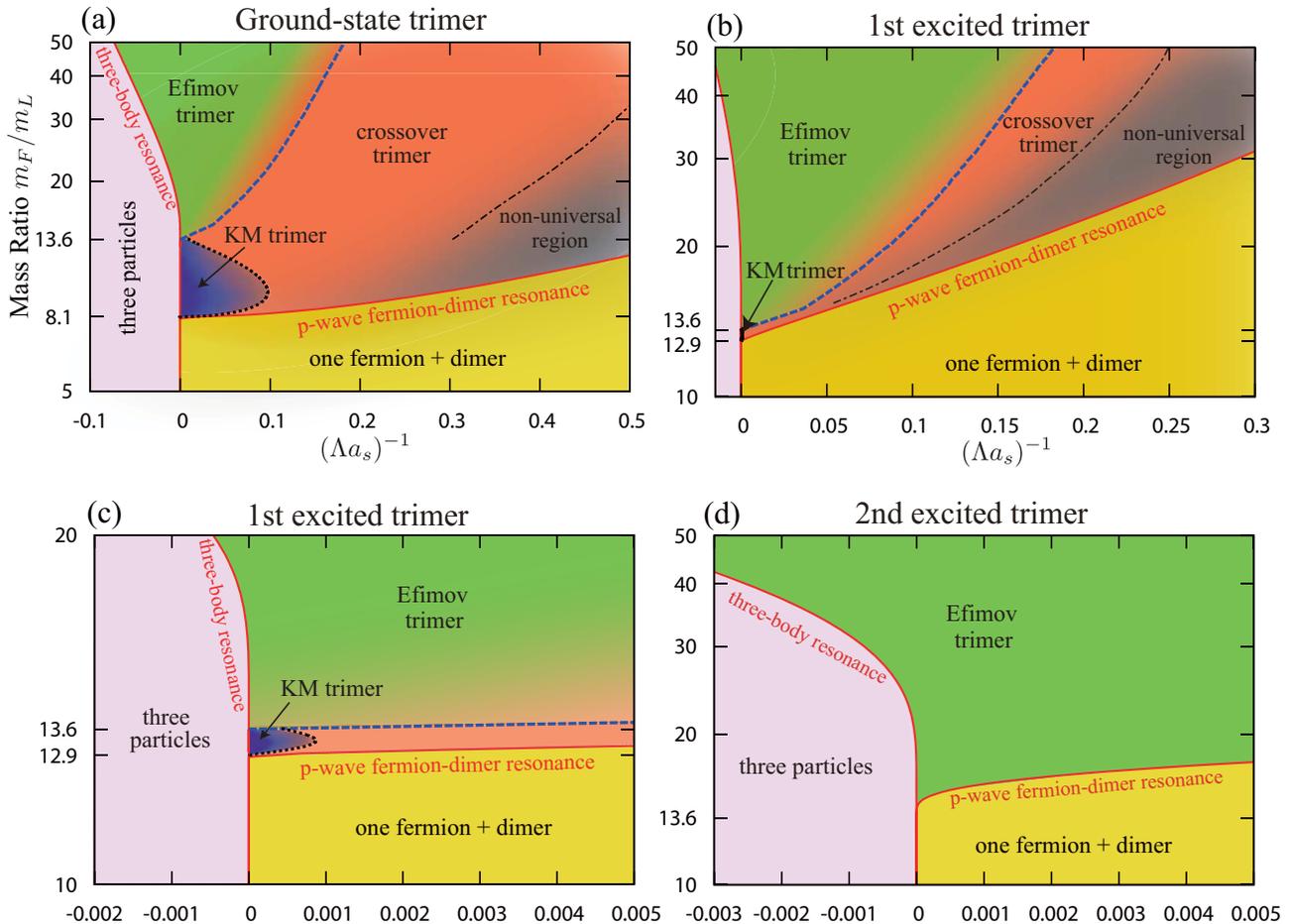}
\caption{\label{fig:All_Schematic_result}(Color online) Regions of (a) the ground-state, (b)(c) first excited, and (d) second excited trimers as a function of the mass ratio and $(\Lambda a_s)^{-1}$, where $\Lambda$ is the three-body parameter and $a_s$ is the $s$-wave scattering length. For the ground-state and first excited trimers, there are four regimes: the Efimov, KM, crossover, and non-universal regimes. For the second excited trimer, only the Efimov region appears. A trimer dissociates into a fermion and a dimer at the fermion-dimer $p$-wave resonance (red curve on the positive $a_s$ side), while it dissociates into three particles at the three-body resonance line (red curve on the negative $a_s$ side). The blue dashed curves and black dotted curves correspond to $q_n = 0.40$ (see \equref{q_n_def}) and $r_n=0.40$ (see \equref{r_n_def}), which delimit the Efimov region and KM trimer region, respectively. The black dashed-dotted curves correspond to $s_n=0.90$ (see \equref{s_n_def}), which separates the crossover and non-universal (i.e., model-dependent) trimer regions. The color contour is used for the sake of clarity. The non-universal region increases with increasing the mass ratio, and the Efimov region disappears for the mass ratio well above 50 (see discussions in \secref{result-model}).}
\end{figure*}

A natural question arises: how does the continuous scaling invariance of KM trimers below the critical mass ratio $(m_F/m_L)_{\mathrm{E}}$ change into the discrete scaling invariance of Efimov trimers above it? Although the two fermions plus one particle system has been studied in some limited regions \cite{efimov1973energy,kartavtsev2007low,PhysRevLett.100.090405,blume2010breakdown,*blume2010few}, there has not yet been a systematic study that varies both the mass ratio and $s$-wave scattering length, and clarifies the relation between Efimov and KM trimers. In this paper, we present a unifying scenario for the Efimov states and KM trimers, which have so far been studied separately. Our findings are summarized as follows:
\begin{enumerate}
\item[1.] There exist trimer states between the Efimov and the KM regimes with neither discrete nor continuous scaling invariance. 
\item[2.] These ``crossover trimer" states appear universally; the crossover trimers appear independently of short-range details of the potential, and characterized only by two parameters, the $s$-wave scattering length $a_s$, and the three-body parameter $\Lambda$. 
\end{enumerate}
We have summarized our main results in Fig.~\ref{fig:Result_Spectrum_Sche} and \ref{fig:All_Schematic_result}. As the mass ratio is varied, we have found that the energy spectrum changes as schematically illustrated in \figref{Result_Spectrum_Sche}. More specifically:
\begin{enumerate}
\item[3.] For $m_F/m_L <(m_F/m_L)_{\mathrm{E}}$, the trimers have the continuous scaling invariance if and only if the $s$-wave scattering length is large, $\Lambda a_s \gg 1$. As we move away from unitarity, they become dependent on the three-body parameter and change into crossover trimers.
\item[4.] At $m_F/m_L =(m_F/m_L)_{\mathrm{E}}$, there exist two trimers on the positive $a_s$ side. These trimers have neither discrete nor continuous scaling invariance and thus they are crossover trimers.
\item[5.] For $m_F/m_L >(m_F/m_L)_{\mathrm{E}}$, the trimers satisfy the discrete scaling law of the Efimov states close to the unitarity point. Away from unitarity, the ground and first-excited trimers deviate from the discrete scaling law and become crossover trimers, while the shallower trimers satisfy the discrete scaling law for the entire region.
\item[6.] Both for $m_F/m_L <(m_F/m_L)_{\mathrm{E}}$ and for $m_F/m_L \ge (m_F/m_L)_{\mathrm{E}}$, the trimers dissociate into a fermion and a dimer on the positive $a_s$ side, and a $p$-wave resonance occurs in the fermion-dimer scattering channel. However, the ground-state trimers may not do so due to non-universal short-range effects.
\item[7.] Close to the fermion-dimer dissociation boundary, the ground-state trimer is model-dependent and hence non-universal. For the first excited trimer, except for a very small region close to the fermion-dimer dissociation boundary, the non-universal short-range effects are negligible. The higher excited trimers are universal (i.e. independent of short-range details) over the entire region.
\end{enumerate}

In \figref{All_Schematic_result}, we present these results as a function of the mass ratio and $1/\Lambda a_s$. By comparing the three-body calculation with a different short-range models, we specify the universal region and non-universal region. Their boundary is shown as black dashed-dotted curves in \figref{All_Schematic_result}. In the universal region, the trimer can be described by the two parameters, while it depends on other short-range details in the non-universal region. Furthermore, in the universal region, we classify the trimers from their scaling properties:
\begin{itemize}
\item Efimov trimer: trimer with a discrete scaling invariance.
\item KM trimer: trimer with a continuous scaling invariance.
\item Crossover trimer: trimer with no scaling invariance.
\end{itemize}
By investigating the scaling properties, we find their boundaries as dashed and dotted curves in \figref{All_Schematic_result}. Note that all these boundaries are crossover boundaries, and thus there is no clear transition at any specific point. Therefore, as the mass ratio is increased at a fixed scattering length $a_s>0$, the KM trimer gradually loses its continuous scaling invariance and turns into a trimer with no scaling symmetry around the critical mass ratio. If the mass ratio is increased further, it starts to acquire the discrete scaling invariance and turns into the Efimov trimer.

In the next section, we review some basic physics of the Efimov trimers and KM trimers, and describe the method we use in this paper. In \secref{result-small}, and \secref{result-large}, we study the scaling properties of the trimers and identify the KM trimer region for $m_F/m_L <(m_F/m_L)_{\mathrm{E}}$ and the Efimov trimer region for $m_F/m_L >(m_F/m_L)_{\mathrm{E}}$, respectively, when the scattering length is finite. We then show the existence of the ``crossover trimers" in \secref{crossover}. In \secref{result-model}, we investigate to what extent trimers are universal (i.e. model-independent). Finally, in \secref{implications}, we discuss experimental implications of our results.


\section{\label{sec:method}Efimov trimers and Kartavtsev-Malykh trimers}
We solve the three-body problem of two identical fermions and one particle with variable mass ratio $\alpha =m_F/m_L$. The interaction between the two fermions can be neglected at low energy, since they can only interact in odd angular momentum channels. On the other hand, the fermion and the third particle can interact in the $s$-wave channel. If the $s$-wave scattering length is much larger than the range of the interaction, it can be modeled by a contact interaction. With this prescription, two angular momenta become good quantum numbers $(\ell_1,\ell_2)$, where $\ell_1$ is the angular momentum between one of the fermion and the third particle, and $\ell_2$ is the angular momentum between the center of mass of these two particles and the other fermion. If $\ell_1 \neq 0$, the fermion and the third particle cannot interact, so that no trimer can be formed. If $\ell_1=0$, we must solve the three-body Schr\"{o}dinger equation by taking the full three-body correlation into account. With the zero-range interaction, the problem reduces to solving the following Skorniakov--Ter-Martirosian equation (STM equation) \cite{stm,endo2011universal}:
\begin{widetext}
\begin{equation}
\begin{split}
\label{eq:stm} \frac{a_{\ell_2}(p)}{a_s} 
+ \frac{(-1)^{\ell_2 } m_{L}}{\mu_{FL}\pi}\int dq\Bigr(\frac{q}{p}\Bigl)^{\ell_2+1} Q_{\ell_2}\Biggr(\frac{\frac{m_{L}}{{2\mu_{FL}}}[-\varepsilon +p^2+q^2]}{pq}\Biggl)& \frac{a_{\ell_2}(q)/a_s}{\sqrt{-\varepsilon+\frac{2\alpha+1}{(\alpha+1)^2}q^2}-\frac{1}{a_s}} \\
&= \frac{(-2)^{\ell_2}(\ell_2 !)^2}{(2\ell_2 +1)!}\frac{\mu_{DF}m_{L}}{\mu_{FL}^2}\Bigr[\frac{2\mu_{FL}}{m_{L}}\frac{1}{(-\varepsilon +p^2)}\Bigl]^{\ell_2+1},
\end{split}
\end{equation}
\end{widetext}
where $p$ is the relative in-coming momentum between the fermion and the dimer, the $\varepsilon = 2\mu_{FL}E$, $\mu_{FL}=\frac{m_F m_L}{m_L + m_F}$ and $\mu_{DF}=\frac{m_F (m_F + m_L)}{m_L + 2m_F}$ are the reduced masses between the fermion and the third particle, and between the dimer and the fermion, respectively, $Q_{\ell}$ denotes the Legendre polynomial of the second kind, and $a_{\ell}(q)$ is the momentum-dependent $\ell$-th wave scattering length. The STM equation was originally proposed to obtain the fermion-dimer scattering length \cite{stm}. In fact, by taking the energy at the dimer threshold $E= -\frac{1}{2\mu_{FL}a_s^2}$ and solving \equref{stm}, the scattering length of the fermion-dimer $\ell$-th partial wave can be obtained as $a_{\ell}(q=0)$ ($s$-wave scattering length for $\ell =0$, and $p$-wave scattering volume for $\ell=1$). The three-body problem is equivalent to the fermion-dimer scattering problem, and thus we can also use it to investigate the properties of trimers with \equref{stm}. The binding energy of the trimer states can be obtained by searching for the value of $\varepsilon$ at which $a_{\ell}$ diverges. This can be done by finding the value of $\varepsilon$ at which the eigenvalue of the left-hand side of \equref{stm} (seen as an operator acting on $a_{\ell}(q)$) vanishes.

If $\ell_2$ is an even integer, no trimer can be formed due to the Pauli principle between identical fermions \cite{kartavtsev2007low}. Thus, the only possible channels for trimers are $\ell_1=0$ with odd $\ell_2$. In this paper, we focus on the most stable channel $(\ell_1,\ell_2)=(0,1)$.

In this channel, it is known that the Thomas collapse occurs for $\alpha = m_F/m_L >(m_F/m_L)_{\mathrm{E}}=13.606...$ \cite{PhysRev.47.903,PhysRevA.67.010703}, and then the STM equation becomes singular, which signals the formation of an Efimov state. To avoid this singularity, a finite momentum cutoff in the integration is introduced:
 \begin{equation}
\int_{0}^{\infty} dq  \rightarrow \int_{0}^{\Lambda} dq .
\end{equation}
 This momentum cutoff amounts to introducing a new short-range scale in the problem; thus the Thomas collapse is avoided and the ground-state energy becomes finite $E_{\mathrm{GS}}\sim \frac{\Lambda^2}{2 \mu_{FL}}$. Therefore, the momentum cutoff may be regarded as a three-body parameter, which fixes the energy scale of the trimers \cite{rem_TBP}. Then, the Efimov states with the discrete scaling invariance are obtained as $E_{n+1} = e^{-2\pi/\gamma}E_n$, $\langle r \rangle_{n+1}= e^{\pi/\gamma}\langle r \rangle_{n}$, where $\gamma$ is determined from the equation \cite{efimov1973energy, kartavtsev2007low}
\begin{equation}
\label{eq:gamma}0=\frac{1+\gamma^2}{\gamma}\tanh \gamma\frac{\pi}{2}-\frac{2}{\sin 2\omega}\frac{\cosh \gamma \omega}{\cosh \gamma \frac{\pi}{2}} + \frac{\sinh \gamma \omega}{\gamma \sin^2 \omega \cosh \gamma \frac{\pi}{2}},
\end{equation}
where
\begin{equation}
\cot \omega = \frac{\sqrt{1+2 \alpha}}{\alpha}.
\end{equation}
The Efimov trimers are thus described by two parameters: the three-body parameter $\Lambda$ and the $s$-wave scattering length $a_s$.

\begin{figure*}[t]
\includegraphics[width=18cm,clip]{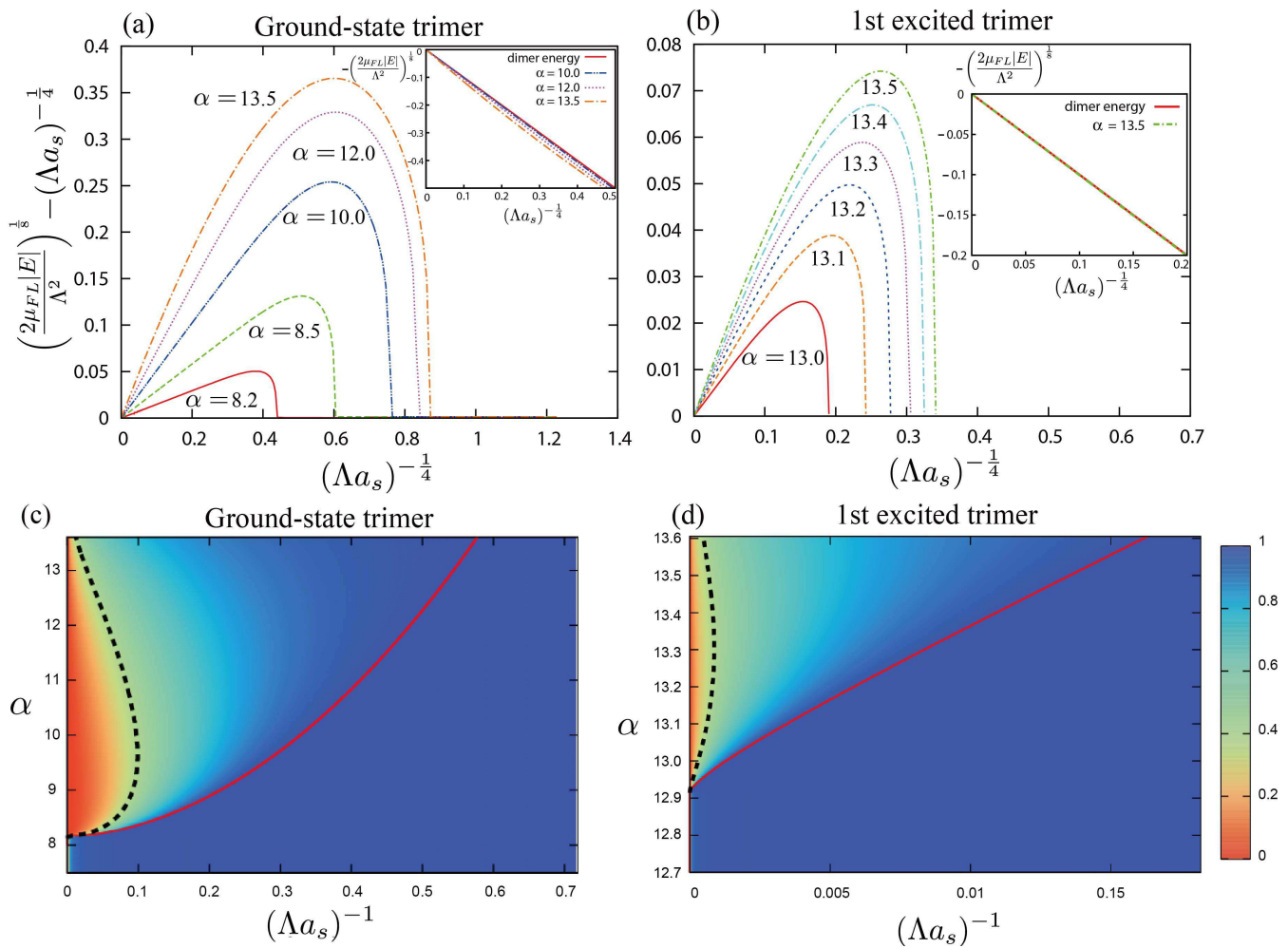}%
\caption{\label{fig:Below_All}(Color online) (a)(b) The spectra of trimers with several different mass ratios $\alpha=m_F/m_L$ as measured from the dimer binding energy for (a) the ground-state trimer and (b) the first excited trimer. The insets show the energy of the trimers as measured from the vacuum, together with the dimer binding energy (red solid curve). (c)(d) Contour plots of $r_n$ defined in \equref{r_n_def} for (c) the ground-state trimer and (d) the first excited trimer. The red solid curve shows the fermion-dimer dissociation threshold, while the black-dotted curve is the $r_n = 0.40$ curve, which delimits the region in which the continuous scaling invariance is well satisfied (i.e., the KM trimer regime).}
\end{figure*}

On the other hand, for $m_F/m_L < (m_F/m_L)_{\mathrm{E}}$, the STM equation is well-behaved at $\Lambda \rightarrow \infty$, and two trimers are known to exist for $a_s>0$ \cite{kartavtsev2007low,endo2011universal}. These trimers start to appear at $(m_F/m_L)_{\mathrm{KM}}^{(1)}=8.172...$ and $(m_F/m_L)_{\mathrm{KM}}^{(2)}=12.917...$, and their binding energy increases with increasing the mass ratio. Since there is only one length scale in the equation, these trimers can be described by the $s$-wave scattering length $a_s$. As a result, these trimers have a continuous scaling invariance: if the $s$-wave scattering length is scaled as $a_s \rightarrow \beta a_s$, the binding energy $E_n$ ($n=1,2$) and the mean radius $\langle r \rangle$ are scaled as $E\rightarrow \beta^{-2}E$, and $\langle r \rangle \rightarrow \beta \langle r \rangle$ for an arbitrary (continuous) value of $\beta$. This is in marked contrast with the discrete scaling invariance of the Efimov states.

The continuous scaling law of the KM trimers is well defined in the scaling limit: $ \Lambda \rightarrow \infty$ with $a_s$ fixed. Thus, to understand the relation between the KM trimers and the Efimov trimers, we must keep $\Lambda$ and $a_s$ finite and investigate how the KM trimers evolve into the Efimov trimers as $\Lambda$, $a_s$, and the mass ratio are varied. Therefore, we solve the STM equation with a finite momentum cutoff $\Lambda$, and examine how well the discrete scaling and continuous scaling hold, thereby presenting a unifying picture for the KM trimers and Efimov trimers.

While we use the STM equation based on the zero-range approximation, this method has been shown to give an accurate description of the three-body system when the scattering length between the particles is large \cite{STM_valid}. Thus, around the unitarity regime, our results should remain unchanged even if one performs more sophisticated three-body calculations. Away from unitarity, the trimer starts to become non-universal and depend on short-range details. In this non-universal region, the STM equation becomes less accurate (see \figref{All_Schematic_result} and discussions in \secref{result-model}).


\section{\label{sec:result}Scaling properties of the trimers}

In this section, we present the results of our numerical calculations with the STM equation for a sharp momentum cutoff. Specifically, we calculate the binding energy of the trimers by changing the mass ratio $\alpha =m_F/m_L$, and the product of the $s$-wave scattering length and the momentum cutoff $a_s \Lambda$. In fact, they are the only two free parameters in \equref{stm} if the energy is measured in units of $\frac{\Lambda^2}{2\mu_{FL}}$. Thus, in this paper, we normalize the energy by $\frac{\Lambda^2}{2\mu_{FL}}$ and investigate the wave number $K\equiv \sqrt{\frac{2\mu_{FL}|E|}{\Lambda^2}}$ as a function of the mass ratio $\alpha$ and the normalized inverse $s$-wave scattering length $(a_s \Lambda)^{-1}$.

\subsection{\label{sec:result-small}$(m_F/m_L)_{\mathrm{KM}}^{(1)}<m_F/m_L<(m_F/m_L)_{\mathrm{E}}$}

The binding energies of the ground-state and first excited trimers are shown in the insets of Figs.~\ref{fig:Below_All}~(a) and (b). The ground-state trimer appears on the positive $a_s$ side at $(m_F/m_L)_{\mathrm{KM}}^{(1)} = 8.172...$, while the first excited trimer appears at $(m_F/m_L)_{\mathrm{KM}}^{(2)} = 12.917...$. The binding energy of the trimers is rather close to the dimer binding energy $\frac{1}{2\mu_{FL}a_s^2}$, suggesting that the three particles are only loosely bound. Note that the trimers do not exist for a negative $a_s$. The trimer binding energies as measured from the dimer binding energy are presented in Figs.~\ref{fig:Below_All}~(a) and (b). A linear behavior can be seen close to the unitarity point. This suggests that the continuous scaling law holds well in this regime and the corresponding trimers are the KM trimers. In fact, the binding energy of the KM trimers should behave as $|E|\propto \frac{1}{2\mu_{FL}a_s^2}$, so that $K^{\frac{1}{4}}\propto (\Lambda a_s)^{-\frac{1}{4}} $, and a linear energy spectrum should be obtained. The existence of the KM trimers for a large positive $a_s$ for $(m_F/m_L)_{\mathrm{KM}}^{(1)}<m_F/m_L<(m_F/m_L)_{\mathrm{E}}$ (ground KM trimer) and $(m_F/m_L)_{\mathrm{KM}}^{(2)}<m_F/m_L<(m_F/m_L)_{\mathrm{E}}$ (first excited KM trimer) was predicted in Ref.~\cite{kartavtsev2007low}, and here the same result is reproduced by a different method.

Away from unitarity, the trimers gradually become dependent on the three-body parameter $\Lambda$. As a result, the deviation from the linear behavior gradually becomes significant, and the trimer finally dissociates into a fermion and a dimer. This suggests that a scattering resonance occurs at this dissociation point in the fermion-dimer scattering channel. Since the trimer we consider here has a finite angular momentum $L_{\mathrm{tot}}=1$, the scattering resonance occurs in the $p$-wave channel. Thus, we arrive at the following conclusions (c.f. statements 3 and 6 in the Introduction):
\begin{itemize}
\item The trimers satisfy the continuous scaling invariance when the $s$-wave scattering length is large $\Lambda a_s \gg 1$.

\item As we move away from unitarity, the continuous scaling invariance deteriorates, and the trimers dissociate into a fermion and a dimer on the positive $a_s$ side as long as non-universal corrections are negligible (see discussions in \secref{result-large} for the non-universal corrections).
\end{itemize}

\begin{figure*}[t]
\includegraphics[width=18cm,clip]{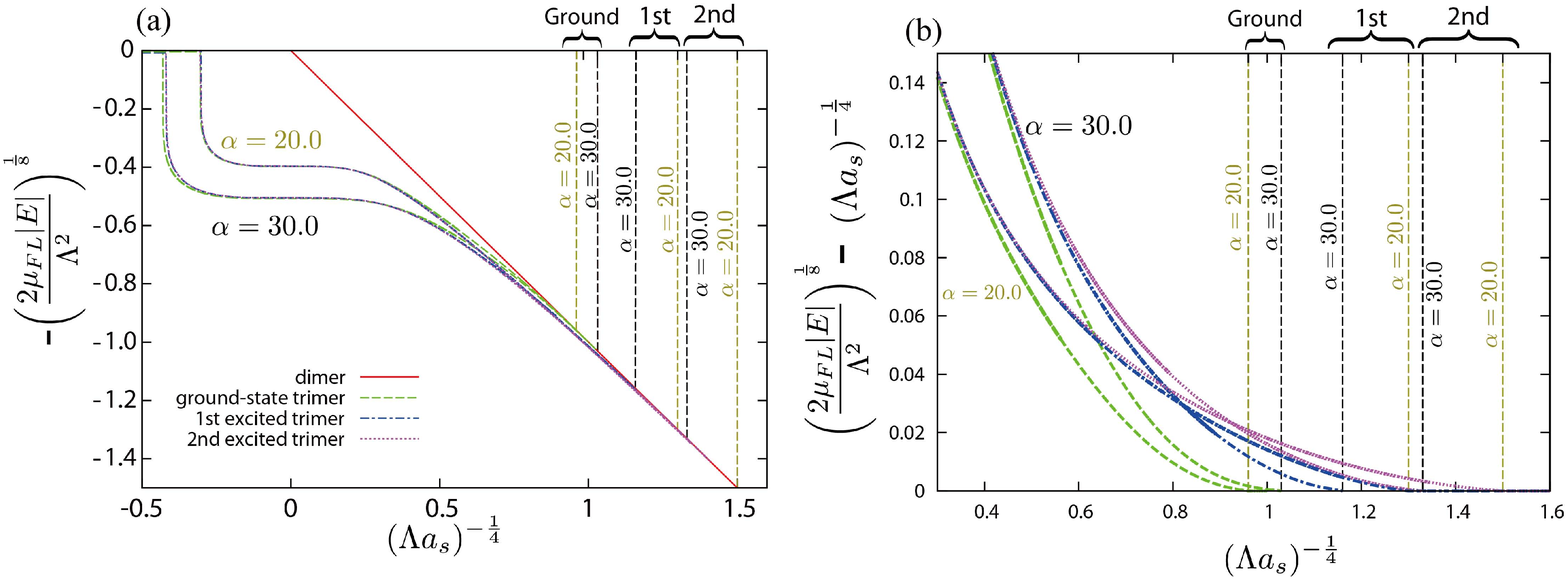}%
\caption{\label{fig:Above_Spectrum_ed}(Color online) (a) Energy spectra of the ground-state, first excited, and second excited trimers for $m_F/m_L =20.0$ and $m_F/m_L =30.0$. The binding energy is measured from the three-body continuum. For the first and second excited trimers, the radial scaling transformation has been performed. On the positive $a_s$ side, the fermion-dimer dissociation points are shown for each trimer state. (b) Energy spectra as measured from the dimer binding energy for $m_F/m_L =20.0$ and $m_F/m_L =30.0$. The same radial scaling transformation is performed as in
 (a).}
\end{figure*} 

One can quantify the continuous scaling property by introducing the following quantity ($x\equiv (\Lambda a_s)^{-1}$):
\begin{equation}
\label{eq:r_n_def}r_{n}(x) = \frac{K_n(x) - K_{\mathrm{dimer}}(x)}{K_n^{(\mathrm{KM})}(x) -K_{\mathrm{dimer}}(x)},
\end{equation}
where $K_{\mathrm{dimer}}(x)= x$ is the dimer binding energy, and $K_n^{(\mathrm{KM})}(x)=\sqrt{ \frac{2\mu_{FL}|E_n|(x)}{\Lambda^2}} = C_n x $ is the trimer binding energy in the limit $1/\Lambda a_s\rightarrow +0$ for the ground-state trimer ($n=1$), and the first excited trimer ($n=2$). $C_n\ge 1$ characterizes the trimer binding energy in the universal limit $\Lambda a_s \gg 1$, and it gets larger as the mass ratio $\alpha$ is increased. Here, $r_n(x)$ is 1 when the continuous scaling law holds exactly, while it is 0 in the absence of the trimers. The values of $r_n$ are shown in Figs.~\ref{fig:Below_All}~(c) and (d). One can clearly see that the continuous scaling law holds sufficiently close to the unitarity limit, and we regard this region as the KM trimer region. Away from unitarity, the trimers do not have the continuous scaling property, and they become dependent on $\Lambda$.

As the mass ratio is increased, the KM trimer region first grows, whereas it starts to shrink as we move closer to the critical mass ratio $(m_F/m_L)_{\mathrm{E}}$. This is a manifestation of the fact that the hyperradial potential barrier at short distance decreases as the mass ratio is increased toward the critical mass ratio (see Figure 1 of Ref. \cite{kartavtsev2007low}). The wave function of the trimers is then more likely to penetrate into a short-distance regime, and the trimers become more sensitive to the three-body parameter $\Lambda$. At the critical mass ratio, the hyperradial potential barrier disappears, so that the trimers depend on $\Lambda$ even close to the unitarity limit, and the continuous scaling law breaks down.

\subsection{\label{sec:result-large}$m_F/m_L>(m_F/m_L)_{\mathrm{E}}=13.606...$}

\begin{figure*}[t]
\includegraphics[width=18cm,clip]{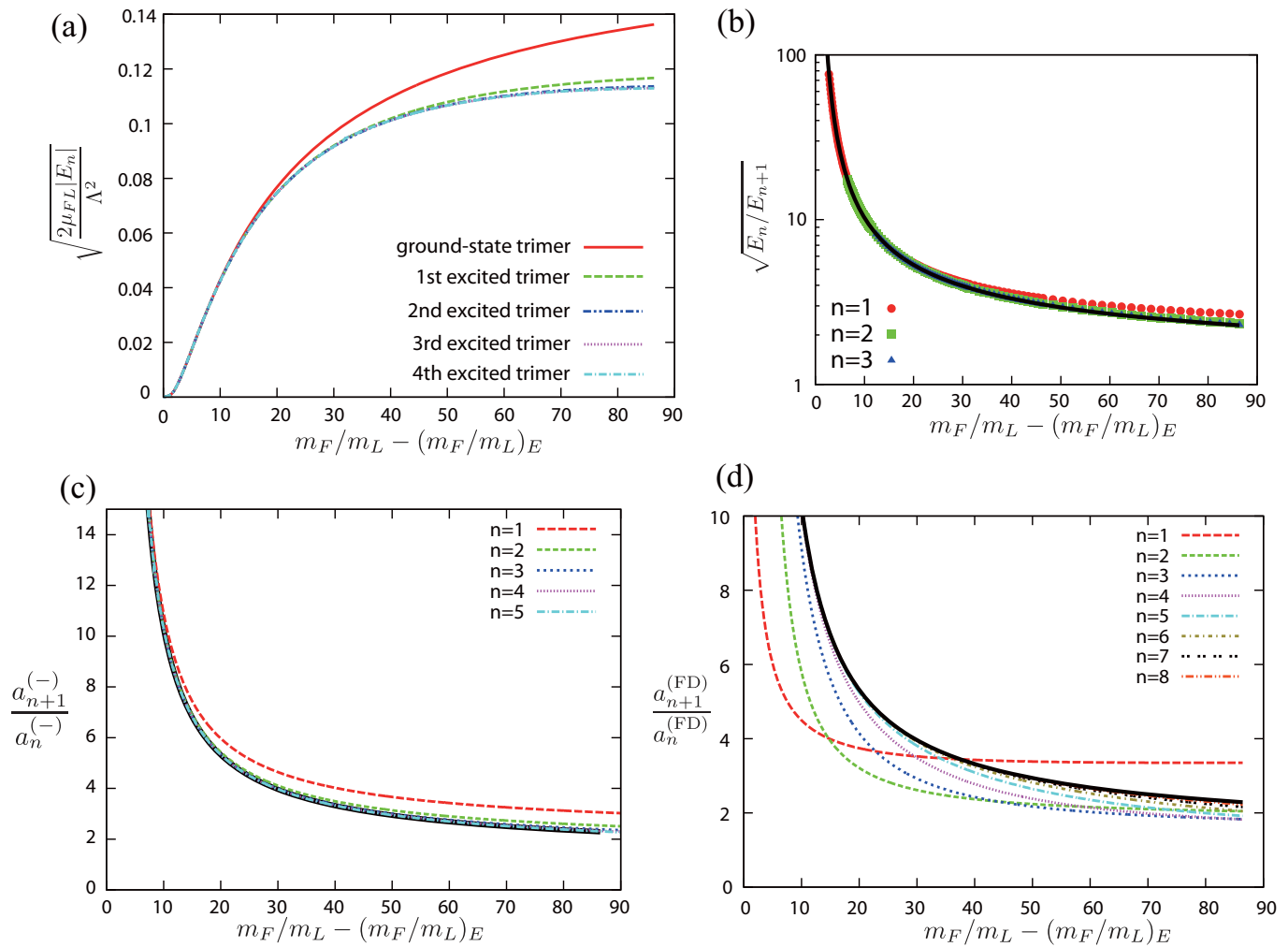}%
\caption{\label{fig:Above_All}(Color online) (a) Binding energy of the ground to fourth excited trimers at the unitarity limit. The radial scaling transformation has been performed. (b) Ratio of the binding energy between the adjacent trimer levels at the unitarity limit. (c) Ratio of the $s$-wave scattering length at which the trimer dissociates into three particles. (d) Ratio of the $s$-wave scattering length at which a trimer dissociates into a fermion and a dimer. In (b)-(d), the universal scaling ratio obtained from \equref{gamma} is shown as black solid curves.}
\end{figure*}

The energy spectra above the Efimov critical mass ratio are shown in Figs.~\ref{fig:Above_Spectrum_ed}~(a) and (b). Here, the following radial scaling transformation is performed for the first and second excited trimers:
\begin{eqnarray*}
\mathrm{first\  excited:}K_2 \rightarrow K_2 e^{\frac{\pi }{\gamma}} \ \ \ & (\Lambda a_s)^{-1} \rightarrow (\Lambda a_s)^{-1}e^{\frac{\pi }{\gamma}}\\
\mathrm{second\  excited:}K_3 \rightarrow K_3 e^{\frac{2\pi }{\gamma}} \ \ \ & (\Lambda a_s)^{-1} \rightarrow (\Lambda a_s)^{-1}e^{\frac{2\pi }{\gamma}},
\end{eqnarray*}
where $\gamma$ is calculated from \equref{gamma}. We perform this scaling transformation so that all the curves should superimpose into a single universal curve if the discrete scaling law of the Efimov states holds. We can see from Fig.~\ref{fig:Above_Spectrum_ed}~(a) that the discrete scaling law holds well for most regions. While the trimers exist only on the positive $a_s$ side for the ground-state and first excited trimers below the critical mass ratio $(m_F/m_L)_{\mathrm{E}}$, above it they exist even for $a_s<0$. In addition to these two trimers, an infinite sequence of trimers start to appear for $m_F/m_L>(m_F/m_L)_{\mathrm{E}}$. As $a_s$ is varied across unitarity, these trimer levels appear on the negative $a_s$ side, and they finally dissociate into a fermion and a dimer on the positive $a_s$ side.

All of these trimer states, including the ground-state and first excited trimers, have a good discrete scaling invariance close to the unitarity point. In \figref{Above_All}~(a), the binding energies from the ground-state up to the fourth excited trimers at the unitarity limit are shown. At the unitarity limit, the binding energies of the trimers evolve smoothly from the critical mass ratio. This behavior is also reported in Ref. \cite{PhysRevA.84.062704} for the same system with a narrow resonance. One can take the ratio $\sqrt{\frac{E_{n}}{E_{n+1}}}$ of the adjacent trimer binding energies, and compare it with a universal discrete scaling factor predicted from the Efimov theory (see \equref{gamma}). The result is shown in  \figref{Above_All}~(b). The discrete scaling law holds for all the trimers including the ground-state and first excited trimers, which are the KM trimers for $m_F/m_L<(m_F/m_L)_{\mathrm{E}}$. Thus, the KM trimers for $m_F/m_L<(m_F/m_L)_{\mathrm{E}}$ and $a_s >0$ change continuously into the Efimov trimers for $m_F/m_L>(m_F/m_L)_{\mathrm{E}}$ and $1/a_s =0$ as the mass ratio and the scattering length are varied. As the mass ratio increases and so does the binding energy of trimers, the Efimov states gradually become dependent on non-universal short-range details and there appear slight deviations from the universal scaling law, as can be seen in Figs.~\ref{fig:Above_All}~(a) and (b).

Figure \ref{fig:Above_Spectrum_ed}~(b) shows that away from unitarity, the binding energy curve becomes less scale invariant. One can quantify this point by observing the position of the three-body threshold $a_n^{(-)}$ and the fermion-dimer dissociation point $a_n^{(\mathrm{FD})}$. In \figref{Above_All}~(c), the ratios of the three-body threshold $a_n^{(-)}$ between adjacent levels are shown. Again, if the discrete scaling law holds, all the curves should be superimposed onto a single universal curve. One can see that the discrete scaling law holds rather well for most of the region. For a large mass ratio, there is a slight deviation from the universal discrete scaling law, especially for small $n$ with a large mass ratio. For these trimers, the binding energy is large, and the non-universal finite-range effect becomes non-negligible, so that the trimers become less Efimovian.

On the other hand, the fermion-dimer dissociation point $a_n^{(\mathrm{FD})}$ behaves quite differently from the three-body threshold as shown in \figref{Above_All}~(d). The ratio of $a_n^{(\mathrm{FD})}$ between the adjacent levels has the following important features:
\begin{itemize}
\item  For $n=1$ (ground - first excited trimers) and $n=2$ (first excited - second excited trimers), the ratio $a_{n+1}^{(\mathrm{FD})}/a_n^{(\mathrm{FD})}$ differs significantly from the universal curve.
\item  For $n\ge 3$, the ratio is consistent with the universal curve if the mass ratio is not too large.
\end{itemize}
The first feature is a consequence of the existence of two KM trimers below the critical mass ratio. In fact, as shown in \figref{Below_All}, the ground-state and first excited trimers exist on the positive $a_s$ side, so that the fermion-dimer dissociation points are well separated from the unitarity point at the critical mass ratio. This is in marked contrast with the fermion-dimer dissociation points for the highly excited trimers (red solid curve for $a_s>0$ in \figref{All_Schematic_result}~(d)), and the three-body thresholds for all the trimers (red solid curve for $a_s<0$ in Figs.~\ref{fig:All_Schematic_result}~(a) to (d)), where they start to appear from the unitarity point at the critical mass ratio. Therefore, the ground-state and first excited trimers no longer satisfy the discrete scaling invariance close to the fermion-dimer dissociation point, while they have a good scaling invariance for $n\ge 3$ as long as the binding energy is small and the non-universal short-range effect is negligible. 

Note here that the fermion-dimer resonance of the ground-state trimer may not appear for some potentials. It is known for the case of three identical bosons that non-universal finite-range effects can prevent the ground-state trimer from dissociating into a fermion and a dimer. This is the case for instance for helium atoms with a scaled realistic potential \cite{PhysRevA.86.012502}. For the two fermions and one particle system we consider here, a similar situation may occur for the ground-state trimer. In contrast, the binding energies of the first excited and higher excited trimers are so small that the finite-range effects should be less significant. Therefore, they are likely to dissociate into a fermion and a dimer.

Thus, we arrive at the following conclusions (c.f. statements 5 and 6 in Introduction):
\begin{itemize}
\item The trimers satisfy the discrete scaling law of the Efimov states close to the unitarity point. Away from the unitarity, the two lowest trimers deviate from the discrete scaling law, while the shallower trimers satisfy the discrete scaling law well for most of the region.
\item The trimers dissociate into a fermion and a dimer on the positive $a_s$ side, and a $p$-wave resonance occurs in the fermion-dimer scattering channel. The ground-state trimers may not do so due to non-universal short-range effects.
\end{itemize}
Again, these statements are valid as long as the mass ratio is not too large, so that non-universal short-range effects are negligible.

We note that the three-particle resonances $a^{(-)}_n$ start to appear right above the critical mass ratio $(m_F/m_L)_{\mathrm{E}}$, and thus there is no three-body resonance for $m_F/m_L<(m_F/m_L)_{\mathrm{E}}$. In general, it is possible that a three-particle resonance start to appear in the region $8.62<m_F/m_L < (m_F/m_L)_{\mathrm{E}}$ \cite{PhysRevLett.100.090405}. However, as pointed out in Ref.~\cite{PhysRevLett.100.090405}, such a three-body resonance for $m_F/m_L < (m_F/m_L)_{\mathrm{E}}$ can appear only when the interaction between the fermions is present and its strength is fine-tuned. In our work, we do not include the interaction between the fermions, since it is negligible in general at low energy. Thus, we do not find such three-body resonance for $m_F/m_L < (m_F/m_L)_{\mathrm{E}}$.


\subsection{\label{sec:crossover}Crossover trimers}
From the discussions so far, the following properties for the ground-state and first excited trimers have been obtained:
\begin{itemize}
\item For $m_F/m_L<(m_F/m_L)_{\mathrm{E}}$, the two trimers satisfy a continuous scaling invariance fairly well if and only if the $s$-wave scattering length is large, $\Lambda a_s\gg 1$, and the continuous scaling region shrinks as the mass ratio is increased (see Figs.~\ref{fig:Below_All}~(c) and (d)).
\item For $m_F/m_L>(m_F/m_L)_{\mathrm{E}}$, the two trimers are Efimov trimers close to the unitarity limit as shown in  Fig.~\ref{fig:Above_All}~(b), whereas they are no longer Efimov trimers close to the fermion-dimer dissociation point (see Fig.~(d)).
\end{itemize}
For $m_F/m_L<(m_F/m_L)_{\mathrm{E}}$, the KM trimer regime is identified by $r_n$. In a similar manner, for $m_F/m_L>(m_F/m_L)_{\mathrm{E}}$, we introduce the following quantity to identify the Efimov trimer region ($x\equiv (\Lambda a_s)^{-1}$):
\begin{equation}
\label{eq:q_n_def}q_n \Bigr(e^{\frac{\pi}{\gamma}}x\Bigl) \equiv \frac{\Bigl|e^{-\frac{\pi}{\gamma}} K_n\Bigr(e^{\frac{\pi}{\gamma}}x\Bigl) - K_{n+1}(x)\Bigl|}{\Bigl|K_{n+1}(x)-x\Bigl|}.
\end{equation}
In \figref{All_Schematic_result}, the contour of $q_n=0.40$ is shown as blue dashed curves. Close to the unitarity point, the Efimov's discrete scaling law holds well, so that $q_n$ is rather small. As we move away from the unitarity, the deviation from Efimov's discrete scaling law becomes significant, and $q_n$ increases. We can identify the Efimov trimer as a region with small $q_n$. As we discussed in \figref{Above_All}~(d), the breakdown of the scaling invariance is significant for the ground-state and the first excited trimers, and the higher excited trimers satisfy the discrete scaling invariance well. If we delimit the Efimov region according to the value of $q_n$, the Efimov region shrinks as the mass ratio is decreased toward the critical mass ratio, as shown in \figref{All_Schematic_result}. Close to the critical mass ratio, both the KM trimer region and the Efimov trimer region shrink, suggesting that the trimers start to acquire a distinct nature from these two trimers around the critical mass ratio; they have neither discrete nor continuous scaling invariance. Thus, we arrive at the following conclusions:
\begin{itemize}
\item The KM trimers for $m_F/m_L<(m_F/m_L)_{\mathrm{E}}$ change continuously into Efimov trimers for $m_F/m_L>(m_F/m_L)_{\mathrm{E}}$ as the mass ratio and the $s$-wave scattering length are varied.
\item In between the KM and Efimov trimers, there exist ``crossover trimers", which have neither discrete nor continuous scaling invariance.
\end{itemize}
The change in the scaling invariance occurs gradually as a crossover, rather than as an abrupt change. The crossover trimer regions are shown as red regions in \figref{All_Schematic_result}. At the critical mass ratio, there is neither the discrete nor continuous scaling invariance. In fact, we can show that there is neither $R^{-2}$ attraction nor a large centrifugal barrier in the hyperradial potential at the critical mass ratio, so that the trimers depend on the three-body parameter $\Lambda$, but still they do not have the discrete scaling invariance of the Efimov states. Thus, we arrive at the first and fourth conclusions listed in Sec. 1.


\section{\label{sec:result-model}Universality: independence on short-range details.}

\begin{figure*}[t]
\includegraphics[width=18cm,clip]{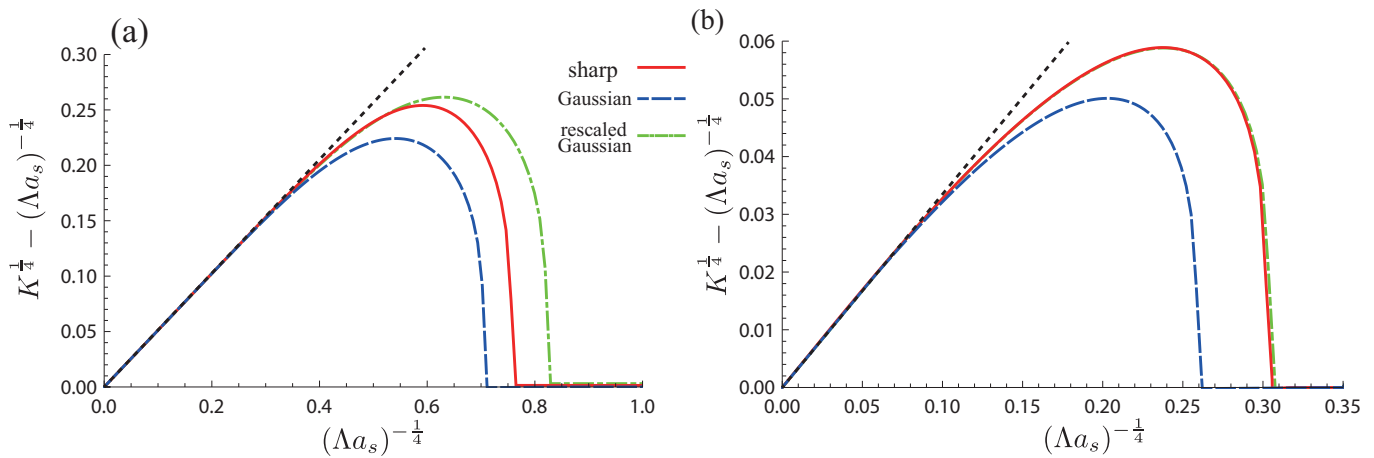}%
\caption{\label{fig:Below_model_comb}(Color online) Energy of the trimers as measured from the dimer binding energy for (a) the ground-state trimer at $m_F/m_L=10.0$ and (b) the first excited trimer at $m_F/m_L=13.3$. The energy for the sharp momentum cutoff (red solid curve) and Gaussian momentum cutoff (blue dashed curve) are compared. The energy of the Gaussian cutoff after a rescaling of the three-body parameter is also shown as green dashed-dotted curves. The scaling factors are taken to be 1.85 for (a) and 1.90 for (b). The black dotted lines are the linear fits in the large $a_s$ region.}
\end{figure*} 

So far, the properties of the trimers by the STM equation with a sharp momentum cutoff have been discussed. This amounts to assuming a certain form of a finite-range potential at short distance and hence imposing a specific value of the three-body parameter. However, one can take other forms of potentials, in general. Thus, one may ask the following question: do the results obtained so far represent universal features of the trimers for the two fermions plus one particle system, or do they represent the special features of the sharp momentum cutoff? The word ``universal" is used here in the sense that the trimers can be characterized only by the three-body parameter and the $s$-wave scattering length. A property is defined to be universal if all the details of the potential other than these two parameters are unimportant. This question can be answered by performing the three-body calculation with different short-range models and comparing the results. In this section, the results of the STM equation calculated with the two different ways of the momentum cutoffs are compared: the sharp cutoff 
 \begin{equation}
\int_{0}^{\infty} dq  \rightarrow \int_{0}^{\Lambda_{\mathrm{S}}} dq, 
\end{equation}
and the Gaussian cutoff
 \begin{equation}
\int_{0}^{\infty} dq  \rightarrow \int_{0}^{\infty} dq \exp\left(-\frac{q^2}{2 \Lambda_{\mathrm{G}}^2}\right).
\end{equation}
If the trimer is universal, the change of the ways of cutoff merely corresponds to changing the value of the three-body parameter. Then, the energy spectra for the two models differ only by their energy scales specified by their three-body parameters. Thus, by rescaling the energy spectra and thereby taking a common three-body parameter, we obtain a common universal spectrum for the two models. In fact, if the energy spectrum $K_{\mathrm{G,S}} = \sqrt{\frac{2\mu_{FL}|E_{\mathrm{G,S}}|}{\Lambda_{\mathrm{G,S}}^2}}$ is plotted as a function of $(\Lambda_{\mathrm{G,S}} a_s)^{-1}$, or $K_{\mathrm{G,S}}^{\frac{1}{4}}$ as a function of $(\Lambda_{\mathrm{G,S}} a_s)^{-\frac{1}{4}}$, then the change in the momentum cutoff corresponds to a mere radial scaling transformation. Thus, whether the trimer is universal or not can be checked by performing a scale transformation of the energy spectra and see whether they can be superimposed upon each other.

 The universal and non-universal regions obtained this way are closely related to the level of approximation of the STM approach. In the universal region, the system can be described by the two parameters, so that the exact three-body calculation should agree with the STM result as far as the scattering length and the three-body parameter are the same. On the other hand, in the non-universal region (shown as gray regions in \figref{Result_Spectrum_Sche}), other short-range details which are not incorporated in the STM approach cannot be ignored. Thus, the STM result will in general deviate from the full quantum calculation with a realistic potential. Our numerical results in this non-universal region are at most qualitative.

Our main conclusions in this section are items 2 and 7 described in the introduction. More specifically, we have found the following:
\begin{itemize}
\item For $m_F/m_L<(m_F/m_L)_{\mathrm{E}}$, the ground-state trimer becomes universal close to the fermion-dimer dissociation point, while the first excited trimer is universal for the most region.
\item For $m_F/m_L>(m_F/m_L)_{\mathrm{E}}$, the ground-state and first excited trimers become non-universal close to the fermion-dimer dissociation point, while they are universal for the other region. 
\item For $m_F/m_L>(m_F/m_L)_{\mathrm{E}}$, the higher excited trimers are universal for the entire region, even close to the fermion-dimer dissociation point.
\item For a mass ratio well above 50, the binding energy of the trimers becomes so large that the non-universal short-range effects can no longer be neglected. 
\end{itemize}
Again, these statements are valid as long as the mass ratio is not too large, so that non-universal short-range effects due to the large binding energy are negligible.

\subsection{$m_F/m_L<(m_F/m_L)_{\mathrm{E}}=13.606...$}

\begin{figure*}[t]
\includegraphics[width=18cm,clip]{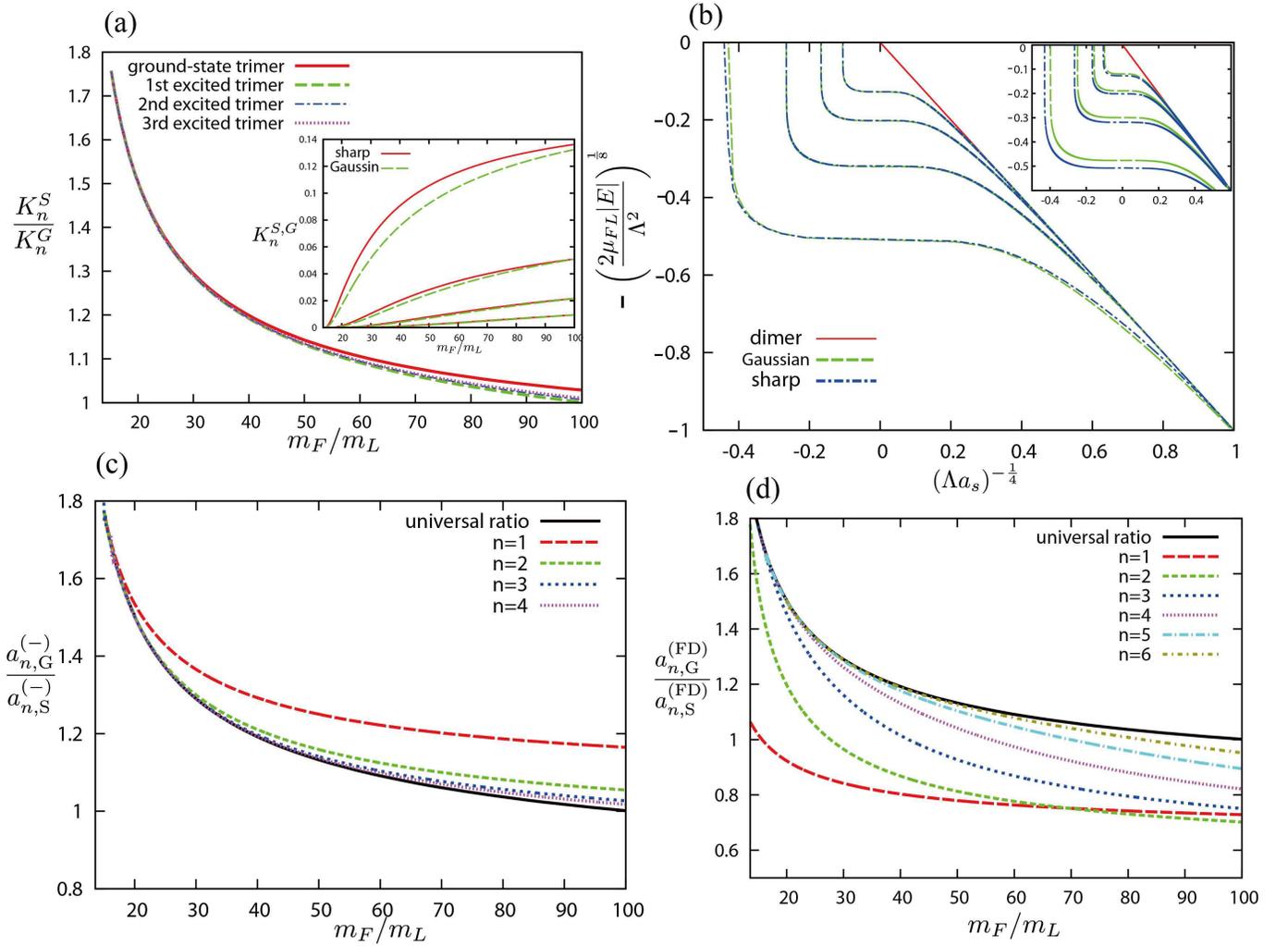}%
\caption{\label{fig:Above_modelIndep_All}(Color online) (a) Ratio of the binding energy between adjacent trimer levels at the unitarity limit. Two models are compared: the STM equation with a sharp momentum cutoff and the Gaussian cutoff. (b) Energy spectra of the ground-state trimer to the fourth excited trimer at $m_F/m_L =30.0$ for the sharp momentum cutoff (blue dashed-dotted curves) and the Gaussian momentum cutoff (green dashed curves). The radial scaling transformation has been performed, so that the energy spectra with the different cutoffs are superimposed onto universal curves. In the inset, the energy spectra before the radial scaling transformation are shown. (c) Ratio of the $s$-wave scattering length at which the trimer dissociates into three particles. (d) Ratio of the $s$-wave scattering length at which the trimer dissociates into a fermion and a dimer. In (c) and (d), the universal scaling curves between the two models obtained from \figref{Above_modelIndep_All}~(a) are shown as the black solid curves.}
\end{figure*} 

The energy spectra as measured from the dimer binding energy are shown in \figref{Below_model_comb}. The energy spectra for the sharp cutoff and Gaussian cutoff agree close to $1/a_s =0$. This is a natural consequence of the nature of the KM trimers: they only depend on the $s$-wave scattering length $a_s$. Since the same $s$-wave scattering length is used for both models, a change in the momentum cutoff does not affect the energy spectrum.  

Away from unitarity, the trimers become dependent on the value of the three-body parameter, and the energy spectra for the sharp and Gaussian cutoffs start to deviate from the linear spectrum of the KM trimers, and the two models give different binding energy curves. However, the two spectra can be superimposed into a single universal curve by performing the rescaling to set a common value of the three-body parameter if the result is model-independent.  In \figref{Below_model_comb}, the energy spectra after rescaling are also shown. For the ground-state trimer, the two curves overlap for $(\Lambda a_s)^{-1/4}\lesssim 0.6$, so that it is model-independent in this region. Close to the fermion-dimer dissociation point the two curves do not overlap, and therefore the ground-state trimer depends on non-universal short-range effects in this region. On the other hand, as we can see in \figref{Below_model_comb}~(b), the first excited trimer is universal for almost the entire region. This difference is closely related to the fact that the binding energy of the first excited trimer is small and thus the $s$-wave scattering length at the fermion-dimer dissociation point is large, compared with the deep ground-state trimer. In fact, $(\Lambda a_s)^{-1}=0.008$ for the first excited trimer at $m_F /m_L =13.3$ with the sharp momentum cutoff. Thus, the dependence on the non-universal short-range details of the models becomes less significant for the first excited trimer. From this fact, we can expect that the fermion-dimer $p$-wave resonance for the ground-state trimer is non-universal, while it is rather universal for the first excited trimer. 

Note that the universal region is larger than the linear region both in \figref{Below_model_comb}~(a) and (b). This fact supports our main conclusion illustrated in \figref{Result_Spectrum_Sche}~(a). For the ground-state trimer, the continuous scaling law holds well for a large $s$-wave scattering length region, and it is the KM trimer. As we vary the $s$-wave scattering length, the ground-state trimer loses its continuous scaling invariance, and this deviation occurs in a model-independent manner. Thus, the KM trimer changes into the crossover timer. As the $s$-wave scattering length is varied further, non-universal short-range effects become significant and the trimer is no longer universal. On the other hand, the first excited trimer is universal for most of the region, so that the non-universal is irrelevant, as described in the upper branch of \figref{Result_Spectrum_Sche}~(a). As the mass ratio is increased toward the critical value, the KM region shrinks, so that the ground-state and first excited trimers become crossover trimers for most of their region, as illustrated in \figref{Result_Spectrum_Sche}~(b).

\subsection{$m_F/m_L>(m_F/m_L)_{\mathrm{E}}=13.606...$}

In Figs.~\ref{fig:Above_modelIndep_All}~(a) and (b), the energy spectra at the unitarity limit and at a fixed mass ratio for the sharp and Gaussian cutoff are shown. As shown in the inset, the energy spectra are different for the two models. As we did for the lower mass-ratio region, the two models should be transformed into each other by rescaling the momentum cutoff as long as the trimers are universal. In \figref{Above_modelIndep_All}~(a), we can see how well this rescaling works at the unitarity limit. At the unitarity, the ratio of the binding energies for the two models gives the rescaling factor, and we can see that the rescaling scenario of the three-body parameter works well for all trimer levels. Therefore, the trimers are model-independent and hence universal. One can see that the universality deteriorates for a large mass ratio, since the binding energy of the trimers is no longer small.

From the rescaling factor between the two models obtained at the unitarity limit shown in \figref{Above_modelIndep_All}~(a), the radial scaling transformation for each mass ratio can be performed, so that we can check whether the two models give the same result after performing the radial rescaling transformation and setting the same three-body parameter. In \figref{Above_modelIndep_All}~(b), the energy spectra after this rescaling are shown. One can see that the two results give the same universal curves for most of the region, so that most of the features of the trimers are model-independent. The disagreement is visible only for regions well separated from the unitarity point.

One can quantify the universality of the three-body threshold and the fermion-dimer dissociation point by taking the ratio of $a_s$'s at which the three-body resonance and the fermion-dimer resonance occur. The results are shown in Figs.~\ref{fig:Above_modelIndep_All}~(c) and (d). The ratio of the three-body thresholds presented in Fig.~\ref{fig:Above_modelIndep_All}~(c) is consistent with the universal curve and thus they are model-independent in most of the region. There are small non-universal deviations only when the mass ratio becomes too large and the binding energy of the trimers becomes large. In Fig.~\ref{fig:Above_modelIndep_All}~(d), on the other hand, there is a significant deviation  for $n=1$ (ground - first excited trimers). This means that the non-universal corrections are significant for the ground-state trimer close to the fermion-dimer dissociation point. For the first excited and higher excited trimers, the deviation is less significant. For these trimers, the binding energy is small, so that they are less affected by the non-universal short-range effects. 

With these results, we can understand the behavior of the trimers described in \figref{Result_Spectrum_Sche}~(c). For the second and higher excited trimers, the spectra satisfy Efimov's discrete scaling law, as discussed in \figref{Above_All}, and they are model-independent for the entire region. For the first excited trimer, the spectrum satisfies Efimov's discrete scaling law for $1/a_s<0$ or close to the unitarity limit. As we change $\Lambda a_s$ toward the fermion-dimer dissociation point, there is a deviation from the discrete scaling law as presented in \figref{Above_All}~(d). This deviation is a universal feature due to the presence of the KM trimers below the critical mass ratio, and it is distinct from non-universal short-range effects. In fact, we can define the following quantity to characterize the model-independence ($x\equiv (\Lambda a_s)^{-1}$)
\begin{equation}
\label{eq:s_n_def}s_n(x) \equiv \frac{\left|K_n^{\mathrm{S}}\left(x \right) - \beta_{\mathrm{SG}}K_{n}^{\mathrm{G}}(\beta_{\mathrm{SG}}^{-1} x)\right|}{\left|K_n^{\mathrm{S}}(x)-x \right|},
\end{equation}
where $\beta_{\mathrm{SG}} \equiv K_n^{\mathrm{S}}/K_n^{\mathrm{G}}$ is a scaling factor between the two models obtained by taking the ratio of the binding energies of the two models at the unitarity (c.f. see \figref{Above_modelIndep_All} (a)). Note that this quantity is well-defined only above the critical mass ratio $m_F/m_L >(m_F/m_L)_{\mathrm{E}}$. Close to the unitarity limit, $s_n$ is small, indicating that the trimers are model-independent. As we move away from unitarity toward the positive $a_s$ side, the non-universal short-range effects becomes significant for the ground-state and first excited trimers and  $s_n$ increases. In Figs.~\ref{fig:All_Schematic_result} (a) and (b), the contour of $s_n=0.90$ is shown as a black dashed-dotted curves, which can be regarded as boundaries between the universal (model-independent) regions and non-universal (model-dependent) regions. While the boundaries for the Efimov trimer regions and the KM trimer regions shrink toward the critical mass ratio, the curves delimiting the universal and non-universal regions are well separated from the unitarity limit. Therefore, the trimers at the critical mass ratio are universal states which are distinct from the KM trimers or Efimov trimers. We can identify them as the crossover trimer states: universal three-body bound states with neither continuous nor discrete scaling invariance. 


The absence of the scaling invariance for the crossover trimer is qualitatively different from the non-universal deviation of the ground Efimov trimer observed in ultracold atom experiments \cite{PhysRevLett.105.023201,ferlaino2011efimov}. This non-universal deviation is due to the large binding energy of the ground-state trimer, and it depends on microscopic details of the potential. On the other hand, we obtain the same binding energy curve of the crossover trimers for different forms of the potentials after the radial rescaling, as we did in Figs.~\ref{fig:Below_model_comb} and \ref{fig:Above_modelIndep_All}~(d). Furthermore, as shown in \figref{Result_Spectrum_Sche}, at the critical mass ratio, the crossover trimers exist even when the scattering length is large $\Lambda a_s \gg 1$. In this limit, the non-universal finite range effects become negligible, and the trimers should behave universally.

\section{\label{sec:implications}Experimental implications}

 To observe the properties of the two fermions and one particle system discussed in this paper, the following conditions are necessary:
 \begin{itemize}
 \item We need a mixture of fermions without internal degree of freedom (i.e. spin-polarized fermions) and another particle. The statistics of the other particle is arbitrary.
 \item The interaction between the fermions and the other particle is resonant, i.e. $a_s \gg \Lambda^{-1} \sim r_0$, where $r_0$ is the range of the interaction.
 \end{itemize}

 Ultracold fermionic gases are the most viable candidates to satisfy these conditions. In ultracold atoms, by using a Feshbach resonance, one can fine-tune the $s$-wave scattering length between the atoms and make it divergently large by changing an external magnetic field \cite{kraemer2006evidence}. Furthermore, one can prepare a mixture of fermionic atoms. For example, a mixture of Li and Yb atoms has been cooled down to quantum degeneracy and an atomic combination with a large mass imbalance is now available \cite{PhysRevLett.106.205304,*hansen2011quantum}. We also note that other atoms with a very large mass are currently being cooled down, such as Dy \cite{PhysRevLett.107.190401}, and Er \cite{aikawa2012bose}. Thus, atomic mixtures with a large mass imbalance seem a promising candidate.

In \tabref{mass_ratio}, the mass ratios for some atomic combinations are presented. Since there appears no trimer for $m_F/m_L<(m_F/m_L)_{\mathrm{KM}}^{(1)}=8.172...$, one must prepare light atoms, such as Li, and heavy atoms, such as Sr, Yb, Er \cite{rem_large_dipole}, and Dy \cite{rem_large_dipole} to observe the trimers. Among those satisfying $m_F/m_L>(m_F/m_L)_{\mathrm{KM}}^{(1)}$, most of them are in the region $m_F/m_L>(m_F/m_L)_{\mathrm{E}}=13.606...$. Thus, both the Efimov trimers and the crossover trimers can be observed with those atomic combinations. On the other hand, there are two candidates, where the KM trimers can be observed: $^6$Li-$^{53}$Cr \cite{rem_large_dipole}, and $^7$Li-$^{87}$Sr. We also note that in ultracold atoms, the effective masses of the atoms can be varied by using an optical lattice \cite{bloch2005ultracold}.
\begin{table}
 \caption{\label{tab:mass_ratio}Mass ratios for some atomic combinations. The two horizontal lines are drawn for the two critical mass ratios $(m_F/m_L)_{\mathrm{KM}}^{(1)}=8.172...$ and $(m_F/m_L)_{\mathrm{E}}=13.606...$.}
 \begin{tabular*}{35mm}{cc}\hline
 \hline
 Species & Mass ratio  \\
\hline
$^7$Li--$^{40}$K  & 5.70 \\
$^7$Li--$^{43}$Ca & 6.12\\
$^{23}$Na--$^{161}$Dy  & 7.00\\
$^{23}$Na--$^{167}$Er  & 7.26\\
$^{23}$Na--$^{173}$Yb & 7.52\\
$^7$Li--$^{53}$Cr &   7.55 \\
\hline
$^6$Li--$^{53}$Cr &   8.80 \\
$^7$Li--$^{87}$Sr &  12.39 \\
\hline
$^6$Li--$^{87}$Sr  & 14.45 \\
$^7$Li--$^{161}$Dy & 22.94\\
$^7$Li--$^{167}$Er & 23.79\\
$^7$Li--$^{173}$Yb & 24.65\\
\hline 
\hline
 \end{tabular*}
 \end{table}

In ultracold atomic gases, the existence of the Efimov states is often identified through the loss of atoms from their confining potential \cite{kraemer2006evidence,ferlaino2011efimov}. Recently, a photoassociation technique is used as an alternative way to observe the Efimov trimers \cite{lompe2010radio,*PhysRevLett.106.143201,*machtey2012association}. This technique has been applied to observe directly the binding energy of Efimov states. With the photoassociation technique, we can measure the binding energy of the trimers as a function of the scattering length, so that we can check how well the continuous or discrete scaling law holds. The deviation from the continuous or discrete scaling invariance as described in \figref{Result_Spectrum_Sche}~(a) and (c) can be a clear signature of the crossover trimers. This universal deviation of the crossover trimer is essentially different from the non-universal corrections discussed for the ground-state Efimov trimer \cite{PhysRevLett.105.023201,ferlaino2011efimov}. While the latter depends significantly on short-range details of each atomic species, the universal deviation can be observed for the crossover trimer after performing the rescaling of the three-body parameter, as we did in \secref{result-model}, and it can be quantitatively predicted with the universal theories, such as the STM equation, or the effective field theory \cite{braaten2006universality}.

\begin{figure}[t]
 \includegraphics[width=9cm,clip]{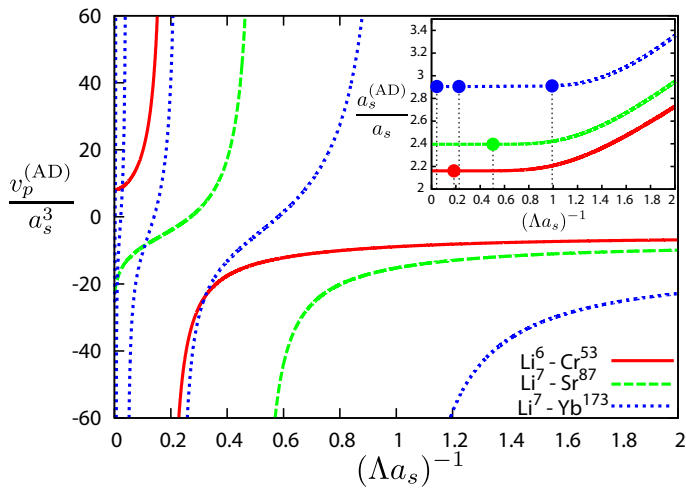}%
 \caption{\label{fig:species_volume_ED}(Color online) $P$-wave scattering volume $v_p^{(\mathrm{AD})}$of the atom-dimer scattering for Li$^6$--Cr$^{53}$ (red solid curve, $m_F/m_L=8.80$), Li$^7$--Sr$^{87}$ (green dashed curve, $m_F/m_L=12.39$), and Li$^7$--Yb$^{173}$ (blue dotted curve, $m_F/m_L=24.65$), calculated with the STM equation with the sharp momentum cutoff. In the inset, the $s$-wave scattering length of the atom-dimer scattering for each atomic combination is shown. The points show the $p$-wave atom-dimer resonance positions for each atomic combination.}
 \end{figure}

Another way is to observe the $p$-wave atom-dimer resonance for $m_F/m_L<(m_F/m_L)_{\mathrm{E}}$. Close to the atom-dimer dissociation point, as shown in \figref{species_volume_ED}, the atom-dimer $p$-wave scattering volume is significantly enhanced, while the atom-dimer $s$-wave scattering length remains of the order of the atom-atom $s$-wave scattering length. This atom-dimer resonance cannot appear for the KM trimers, and therefore it suggests the existence of the crossover trimer, as illustrated in \figref{Result_Spectrum_Sche}~(a). Especially, the $p$-wave atom-dimer resonance for the first excited trimer is a universal feature of the crossover trimer. The enhanced $p$-wave interaction would deform the shape of the cloud, so that it should have a measurable effect on the in-situ and time-of-flight images of the condensates. In addition, one can also measure the enhanced $p$-wave scattering volume by colliding two condensates of atoms and dimers \cite{PhysRevLett.93.173201,*PhysRevLett.93.173202}. Other methods to observe the collisional properties are reviewed in Ref.~\cite{RevModPhys.71.1}.

  Other candidates to observe the KM trimers and crossover trimers are nuclear systems \cite{PhysRevA.34.4424,PhysRevLett.73.2817,nakamura2006observation,nakamura2009halo,AnnRev_HamPlatt}. In some nuclear systems, the $s$-wave scattering length between nucleons or nuclei may be accidentally large, and the possibility to observe Efimov states has been discussed in Ref. \cite{PhysRevA.34.4424}. Recently, neutron halo states are studied in neutron-excess nuclei, especially for heavy nuclei such as $^{11}$Li \cite{nakamura2006observation}, $^{31}$Ne \cite{nakamura2009halo}, and their analogy with the Efimov states has often been discussed \cite{PhysRevLett.73.2817,AnnRev_HamPlatt}. In these systems, the scattering length between the nuclei and neutrons are accidentally large, so that a rather shallow and large trimers may be formed. If there exist some nuclei with a large $s$-wave scattering length and a suitable mass ratio, the universal trimers and crossover trimers may also be observed. We note, however, that the existence of the Coulomb repulsion and the interaction in the higher-angular-momentum channels often significantly affect the halo states in the nuclei, so that these systems may be more complicated than what has been presented in this paper.

\section{Conclusion}
In conclusion, we have found that there exist trimers with neither discrete nor continuous scaling invariance between the KM trimer region with the continuous scaling invariance and the Efimov trimer region with the discrete scaling invariance. These ``crossover trimers" appear regardless of the short-range details of the potential, and they are characterized only by two parameters, the $s$-wave scattering length $a_s$, and the three-body parameter $\Lambda$. As the $s$-wave scattering length and the mass ratio are varied, the KM trimers change continuously into the Efimov trimers via the crossover trimers. Furthermore, we have shown that the energy spectrum changes as described in \figref{Result_Spectrum_Sche} as the mass ratio is varied (see statements 3 to 7 in the Introduction). We have specified the scaling property and universality (i.e., independence of short-range details) of the trimers as shown in \figref{All_Schematic_result}. 

The Efimov trimers and KM trimers can also exist with a higher angular momentum $L_{\mathrm{tot}}$, specifically for two fermions plus one particle systems with odd integer $L_{\mathrm{tot}}$, as well as two bosons plus one particle systems with non-zero even integer $L_{\mathrm{tot}}$ \cite{kartavtsev2008universal,endo2011universal}. Therefore, we expect a scenario similar to the one discussed in this paper for these systems as well. We also note that a change in dimensionality would make this qualitatively different \cite{kartavtsev2009bound,PhysRevLett.103.153202,*PhysRevA.82.033625}, which could be a subject of future study. 

\begin{acknowledgments}
We thank Yusuke Nishida for useful discussions. This work was supported by
KAKENHI 22340114, 
a Grant-in-Aid for Scientific Research on Innovation Areas ``Topological Quantum Phenomena" (KAKENHI 22103005),
the Global COE Program ``the Physical Sciences Frontier,"
and the Photon Frontier Network Program,
from MEXT of Japan. S. E. acknowledges support from JSPS (Grant No. 237049).
\end{acknowledgments}


\begin{thebibliography}{43}
\bibitem [{STM()}]{STM_valid}%
  \BibitemOpen
  \href@noop {} {}\bibinfo {note} {In \cite{PhysRevA.86.012502}, the STM
  equation is compared with a full quantum three-body calculation with a
  realistic helium potential, and an excellent agreement is
  obtained}\BibitemShut {NoStop}%
\bibitem [{rem({\natexlab{b}})}]{rem_large_dipole}%
  \BibitemOpen
  \href@noop {} {} \bibinfo {note} {Cr, Dy, and Er have a
  large magnetic dipole momentum. In Ref.
  \cite{PhysRevLett.106.233201,*PhysRevLett.107.233201}, the Efimov problem
  under the strong dipole interaction is considered. It is shown that there
  appears Efimov states when $|a_s|\gg d_l$, where $d_l$ is the dipole length,
  and the main difference from the system with the contact interaction is that
  the three-body parameter is universally determined $\Lambda^{-1} \sim d_l$.
  Thus, we believe that our conclusion remains the same even under the strong
  dipole interaction if we think $\Lambda^{-1} \sim d_l$.}\BibitemShut {Stop}%
\end{thebibliography}
%

\end{document}